\documentclass[manuscript]{acmart}
\usepackage{graphicx}
\usepackage{booktabs}
\usepackage{tikz}
\usetikzlibrary{arrows.meta,positioning}
\usepackage{iftex}
\ifXeTeX
  \usepackage{xeCJK}
  \IfFontExistsTF{Noto Serif CJK SC}{\setCJKmainfont{Noto Serif CJK SC}}{%
    \IfFontExistsTF{FandolSong-Regular}{\setCJKmainfont{FandolSong-Regular}}{}}
  \newenvironment{CJK}[2]{}{}
  \newcommand{\zh}[1]{#1}
\else
  \usepackage{CJKutf8}
  \newcommand{\zh}[1]{\begin{CJK}{UTF8}{gbsn}#1\end{CJK}}
\fi
\setcopyright{none}
\acmConference[GROUP'27]{ACM GROUP Conference}{2027}{Georgia, USA}

\author{Yibo Meng}
\authornote{These authors contributed equally to this work.}
\affiliation{
  \institution{Cornell University}
  \city{New York}
  \country{USA}
}
\email{mengyb22@mails.tsinghua.org.cn}

\author{Lyumanshan Ye}
\authornotemark[1]
\affiliation{
  \institution{Shanghai Jiao Tong University}
  \city{Shanghai}
  \country{China}
}

\author{Bingyi Liu}
\affiliation{
  \institution{University of Michigan}
  \city{Ann Arbor}
  \state{Michigan}
  \country{United States}
}

\author{Huidi Lu}
\affiliation{
  \institution{Saïd Business School, University of Oxford}
  \city{Oxford}
  \country{United Kingdom}
}
\email{huidi.lu@sbs.ox.ac.uk}
\correspondingauthor

\author{Xiaolan Ding}
\affiliation{
  \institution{North China University of Science and Technology Health Science Center}
  \city{Tangshan}
  \country{China}
}
\correspondingauthor

\begin{document}
\title[Navigating Deception on Gay Dating Apps]{Navigating Deception on Gay Dating Apps: A Qualitative Study of Users' Experiences, Verification Strategies, and Community Knowledge in China}

\begin{abstract}
Gay dating applications have become essential platforms for sexual minority men to seek relationships, intimacy, and community, yet they also expose users to a range of deceptive interactions that remain underexplored in HCI and CSCW research. This study examines how gay male users in China experience, identify, and respond to deception on dating applications. Through semi-structured interviews with 22 participants who used platforms such as Blued, Aloha, Fanka, and Soul, we identify a typology of deceptive practices that users encounter, including relational misrepresentation for sexual access, sustained emotional deception, financial fraud, marketing behavior disguised as ordinary users, and false identity construction. We further analyze the verification and coping strategies that users develop through long-term platform use, such as judging relational intent through interactional pace, identifying financial risk signals, recognizing commercial promotion accounts, and progressively verifying identity authenticity. Finally, we show that risk recognition is not merely an individual capacity but a collaborative practice that gradually emerges through the circulation of experience within the community, the summarization of recurrent deceptive tactics, and the learning of shared rules. Our findings contribute to CSCW scholarship by demonstrating how users of a marginalized community collectively construct practical knowledge about deception, and by revealing that trust assessment on dating platforms operates as a layered, probabilistic, and socially distributed process rather than a binary judgment.
\end{abstract}
\maketitle

\section{Introduction}

Dating applications have transformed how people initiate romantic and sexual relationships, offering unprecedented opportunities for connection across geographic and social boundaries \cite{ref19,ref21}. For gay men, these platforms hold particular significance: in societies where homosexuality remains stigmatized, dating apps such as Grindr, Blued, and Aloha provide relatively safe spaces for identity expression, partner seeking, and community building \cite{ref4,ref53,ref54}. In China, where homosexuality was decriminalized in 1997 and removed from the classification of mental disorders in 2001 but social acceptance remains limited \cite{ref30,ref48}, gay dating applications serve as critical infrastructure for a population with few alternative channels for meeting potential partners. In this paper, we build an empirically grounded typology of how gay male users themselves define and experience deception on these platforms, and we trace how users assess credibility and collectively transform individual experiences into shared knowledge. Concretely, we ask: (RQ1) What different kinds of deceptive interactions do users encounter on gay dating applications? (RQ2) How do users assess the credibility of others and develop verification and coping strategies during platform-based interaction? And (RQ3) how are experiences of deception circulated among users and gradually transformed into shared judgment knowledge within the community? Stating these analytic aims at the outset, we now situate them within prior work on deception and online dating.

However, the same affordances that make dating platforms valuable---low-cost communication, selective self-presentation, and access to a large pool of strangers---also create fertile ground for deception \cite{ref27,ref45,ref47}. Research on deception in online dating has documented how users misrepresent physical attributes, fabricate biographical details, and strategically manage impressions to appear more attractive \cite{ref18,ref43,ref44}. Studies of romance scams have further shown how fraudsters exploit the relational scripts of online dating to extract financial resources from victims \cite{ref49,ref50,ref51}. Yet the majority of this work has focused on heterosexual dating contexts in Western societies, leaving significant gaps in our understanding of how deception operates on platforms primarily used by sexual minority men, particularly in non-Western contexts where the social and regulatory landscape differs markedly.

For gay men in China, the ecology of deception is shaped by distinctive structural conditions. The semi-closeted status of many users amplifies vulnerability to exploitation, as the social costs of exposure can be leveraged by bad actors. Apps such as Blued and Aloha coexist with more general social platforms, creating a complex ecosystem in which users navigate different norms, audiences, and risks \cite{ref52,ref55}. Moreover, the forms of deception users encounter extend well beyond the profile-level misrepresentation that has dominated the literature: relational misrepresentation aimed at obtaining sex, sustained emotional manipulation, financially motivated fraud, and commercial marketing disguised as ordinary interaction---a range that existing frameworks do not adequately capture.

Equally important, existing research has paid relatively little attention to how users actively respond to deception. Studies of online dating have documented self-presentation strategies \cite{ref19,ref21} and uncertainty reduction behaviors \cite{ref14,ref22}, but have less often examined the practical verification and coping strategies that users develop through accumulated platform experience. Moreover, the role of community in shaping individual risk recognition has been largely overlooked. HCI and CSCW research on LGBTQ+ communities has examined identity management \cite{ref16,ref25}, safety \cite{ref9,ref39}, and platform experiences \cite{ref41,ref42}, but has not systematically investigated how users collectively construct and circulate knowledge about deception.

This study addresses these gaps through a qualitative investigation of gay male users' experiences of deception on dating applications in China. Drawing on semi-structured interviews with 22 participants, we ask three research questions, introduced above and restated here for reference:
\begin{itemize}
  \item RQ1. What different kinds of deceptive interactions do users encounter on gay dating applications?
  \item RQ2. How do users assess the credibility of others and develop verification and coping strategies during platform-based interaction?
  \item RQ3. How are experiences of deception circulated among users and gradually transformed into shared judgment knowledge within the community?
\end{itemize}

Our findings make three contributions. First, we provide a grounded typology of deception on gay dating platforms that extends beyond profile-level misrepresentation to encompass relational, emotional, financial, and commercial forms of deception. Second, we document the layered, probabilistic verification strategies that users develop through long-term platform use, showing that trust assessment operates as a gradual, multi-signal process rather than a binary judgment. Third, we demonstrate that risk recognition is a collaborative practice: users share experiences, summarize recurrent tactics, and collectively construct practical rules that shape how newcomers learn to navigate deception. These findings contribute to CSCW scholarship on trust, deception, and collaborative safety practices in platform-mediated interaction.

\section{Related Work}

\subsection{Deception and Self-Presentation in Online Dating}

Before reviewing this literature, we clarify what we mean by deception, since the term is easily conflated with adjacent phenomena. Following Buller and Burgoon \cite{ND-buller1996}, we define deception as a communicator's deliberate attempt to foster in another person a belief or understanding that the communicator considers false. Three features of this definition do analytic work for us. First, deception requires \emph{intent}: an honest mistake or an aspirational self-image is not deception. Second, it requires a target whom the communicator seeks to mislead, which distinguishes deception (a property of an actor's conduct) from \emph{suspicion} (a property of a perceiver's interpretive state); a person may be suspicious without anyone deceiving them, and a deceiver may succeed without ever arousing suspicion. Third, the falsehood concerns a belief the deceiver wishes the target to hold and act upon. This lets us separate deception from three neighbouring concepts. \emph{Selective self-presentation}---disclosing flattering information while withholding the unflattering---and the broader practice of \emph{impression management} \cite{ref24} are not, in themselves, deceptive: an actor who omits or emphasises true facts to shape how they are seen is not trying to instil a belief they hold to be false. Deception begins where such curation crosses into asserting or implying what the actor knows to be untrue, or into cultivating relational beliefs (about intent, affection, or commitment) that the actor does not share. \emph{Scams}, finally, are a narrower subset: instrumentally motivated deceptions, typically organised and financially extractive, in which the relational frame is itself a pretext. Our concern in this paper is broader than scams and distinct from ordinary impression management; it is the deliberate cultivation of false relational beliefs over the course of an interaction.

This framing also requires us to revisit how Goffman has been mobilised in this area. Beyond the dramaturgical model of front-stage performance, Goffman's account of self-presentation is fundamentally \emph{relational}: the self is not a fixed possession that one either honestly discloses or falsifies, but an effect produced collaboratively in interaction, sustained by the ``working consensus'' that participants tacitly negotiate and by the audience's willingness to honour the line a performer projects \cite{ref24}. Read this way, performance is constitutive of, rather than opposed to, sincere self-presentation, and the relevant question is not merely whether a profile is true but whether the relational definition of the situation a performer fosters is one they themselves accept. Deception, on this view, is a breach of the interactional contract through which a credible self is jointly maintained---a framing that motivates our processual, relational treatment below.

Research on deception in online dating has established that misrepresentation is common but typically small in magnitude. Hancock, Toma, and Ellison \cite{ref27} found that deviations in height, weight, and age were frequent but minor. Subsequent work examined how physical attractiveness influences deceptive self-presentation \cite{ref43}, identified linguistic markers of deception in dating profiles \cite{ref44}, explored how users conceptualize profile veracity \cite{ref18}, and linked attachment styles to online dating deception \cite{ref37}. Ellison, Heino, and Gibbs \cite{ref19} documented how daters manage the tension between impression management and authentic self-presentation, while Gibbs et al. \cite{ref21} showed that anticipated future interaction shapes self-disclosure.

Theoretical frameworks have drawn on Goffman's \cite{ref24} dramaturgical model, Walther's \cite{ref46,ref47} hyperpersonal model, and Donath's \cite{ref17} analysis of identity and deception in virtual communities. These frameworks emphasize that online environments afford greater control over self-presentation, enabling both strategic identity construction and deliberate misrepresentation. Ranzini and Lutz \cite{ref38} demonstrated how self-presentation on Tinder is shaped by the interplay of motives and platform affordances.

However, the literature on online dating deception has predominantly focused on profile-level misrepresentation---that is, discrepancies between what users claim about themselves and what is objectively true \cite{ref45}. This focus, while important, captures only one dimension of deception in platform-mediated interaction. Our study extends this work by examining deception as a processual, relational phenomenon that unfolds through interaction over time, encompassing not only false self-presentation but also the manipulation of relational intent, emotional investment, and financial trust.

\subsection{Romance Scams and Financial Fraud Online}

A growing body of research has examined romance scams as a distinctive form of online fraud. Whitty \cite{ref49} provided an anatomy of the online dating romance scam, showing how scammers build trust through sustained communication before introducing financial requests. Whitty and Buchanan \cite{ref50} characterized romance scams as a serious cybercrime and later documented the psychological impact on victims, finding that many experienced the loss of the relationship as more distressing than the financial loss \cite{ref51}. Buchanan and Whitty \cite{ref8} further investigated the causes and consequences of romance scam victimhood, identifying personality traits and online behaviors that increase vulnerability. More recently, Bilz, Shepherd, and Johnson \cite{ref3} conducted a systematic literature review of online romance scam research, mapping the field's evolution and identifying gaps in understanding victim experiences across diverse populations.

Beyond this foundational work, a substantial and rapidly expanding literature has theorized romance scams, identity deception, and online dating fraud more broadly. Cross has examined how romance fraud unfolds and is responded to, how victims come to recognize it, and the disbelief, stigma, and blame they encounter \cite{ND-cross2021,ND-cross2015,ND-cross2023}, while Cross and Layt analyze the gendered and moralizing media framing through which ``you the idiot'' narratives are constructed around victims \cite{ND-crosslayt2022}. Amirkhani and colleagues extend this literature beyond financially motivated fraud, introducing the notion of ``body scams''---romance scams that pursue sexual access rather than money---and tracing the layered victimization that such relational deception produces, including most recently how scammers reach victims by exploiting family and marriage pressure \cite{ND-amirkhani2024,ND-amirkhani2025,ND-amirkhani2026}. Complementing this fraud-centered work, HCI scholarship on identity and safety in online dating---Zytko and Furlo's analysis of consent and self-presentation in dating apps \cite{ND-zytkofurlo2023}, and Vitak et al.'s study of how users navigate privacy, trust, and disclosure with strangers online \cite{ND-vitak2018}---shows that deception is entangled with everyday practices of identity management rather than confined to discrete criminal events.

This literature has important implications for understanding deception on gay dating platforms. However, existing studies of romance scams have primarily examined cases where fraud is the primary and clearly identifiable goal, typically within heterosexual dating contexts. We emphasize that the gap our study addresses is not an absence of scam research as such---which, as the work above attests, is now extensive---but rather the near-total neglect of the gay-male context within it. In our data, financial extraction often occurs within a more ambiguous relational context, where the boundary between genuine interaction and fraud is less clear-cut. Moreover, the forms of deception our participants described extend beyond financial fraud to include relational and emotional manipulation that may not involve any monetary dimension. Our study thus complements the romance scam literature by examining a broader ecology of deception in which financial fraud is one element among several.

\subsection{LGBTQ+ Communities, Dating Platforms, and Safety}

HCI and CSCW research has increasingly attended to LGBTQ+ people's experiences on digital platforms, with work examining identity management \cite{ref16,ref25,ref35}, technology-mediated safety and harm \cite{ref39}, navigation of hostile online environments \cite{ref31,ref32}, and the role of digital spaces in community building \cite{ref28,ref33}. A recent review of queer HCI by Taylor et al. \cite{ref42} synthesizes this trajectory, arguing that scholarship has moved from treating LGBTQ+ users as a population to be studied toward foregrounding how queer experiences reshape core assumptions about safety, visibility, and design. Research on stigma management shows that LGBTQ+ individuals develop complex strategies for identity disclosure online \cite{ref7,ref10}, which shapes their vulnerability to exploitation on dating platforms. Beyond gay men, work on sexual-minority women in China by Cui et al. \cite{ND-cui2022} shows how online dating is negotiated under compounded heteronormative and gendered surveillance, where users calibrate disclosure and verification against the risk of exposure to family and broader social networks.

Research on gay dating applications has examined self-presentation on Grindr \cite{ref4}, uncertainty management in mobile dating \cite{ref14}, and how men-seeking-men platforms shape self-presentation affordances \cite{ref20}. Giles et al. \cite{ref23} and Colliver \cite{ref12} examined safety, risk, and victimization on gay dating apps. Park et al. \cite{ND-park2024} document how queer women in a closeted society navigate the normalization pressures of online dating, showing that judgments about safety and disclosure are conditioned by an unaccommodating social environment, and Park, DeVine, and Toombs \cite{ND-park2025} foreground the reflexive practice that queer HCI research from non-Western perspectives demands---together underscoring that queer platform experiences, and their study, are situated rather than universal. In the Chinese context, Wu and colleagues have documented relationship development on dating apps \cite{ref53}, the domestication of Blued across generational and class lines \cite{ref54}, the construction of sexual fields across pluralized platforms \cite{ref52}, and the intersection of sex, risk, and stigma \cite{ref55}. Miao and Chan \cite{ref36} provided a social constructivist account of Blued, and Zheng and Lin \cite{ref56} analyzed self-presentation patterns revealing appearance focus and exclusion. Safety concerns across LGBTQ+ dating platforms have been documented across harassment \cite{ref40}, privacy risks, and platform governance \cite{ref2,ref41}.

These situated conditions do more than provide a backdrop; they help explain \emph{why} the particular verification and coping strategies we document arise, and why they take a collective form. First, because homosexuality in China is neither criminalized nor institutionally protected, formal recourse is unattractive: reporting deception to platforms or authorities can itself risk disclosure of one's sexuality, so users channel protective effort into informal, low-visibility tactics instead \cite{ND-kong2016,ref55}. Second, the thinness of offline queer infrastructure makes dating platforms close to obligatory passage points; exiting the platform to escape deception is rarely a realistic option, which pushes users toward strategies of \emph{staying while verifying}---slowing interactions, triangulating profiles across platforms, and testing consistency over time \cite{ref53,ref54}. Third, the pervasive risk of exposure to family and acquaintance networks means that verification moves that would be routine elsewhere (video calls, meeting in public, asking mutual contacts) themselves carry disclosure costs, favoring cues that can be read at a distance and tested without revealing oneself \cite{ND-cui2022}. Finally, because these constraints are shared across the community rather than idiosyncratic, individually accumulated lessons acquire collective value: the community-level circulation of deception scripts (\emph{taolu}, \zh{套路}---literally ``set routines,'' recurring scripted tactics that can be named, taught, and recognized) that we analyze in Section~4.3 can be read as a structural adaptation---a distributed verification infrastructure built precisely where institutional verification is unavailable or unsafe \cite{ref33,ref42}.

Despite this growing body of work, the specific question of how gay men experience, recognize, and collectively respond to deception on dating platforms has received limited attention. Existing studies tend to treat safety as a matter of individual risk management rather than as a collaborative, community-level practice. Even recent accounts that emphasize the relational character of safety \cite{ND-park2024,ref42} stop short of examining how deception encounters specifically are circulated and interpreted within a community. Our study addresses this gap by examining how users develop verification strategies not only through individual experience but also through the circulation and collective interpretation of deception encounters within the community.

\subsection{Trust, Uncertainty, and Verification in Online Interaction}

Trust and uncertainty are central concerns in online interaction. Blumenthal and Clark \cite{ND-blumenthal2001} argued that trust on the Internet is not a binary property of systems but an emergent, context-dependent assessment that users continually negotiate, contending that as networked applications proliferate, trust management is best located with end users and the social practices surrounding them rather than encoded solely in technical infrastructure. This perspective frames the trust work we observe on dating platforms as an ongoing, situated accomplishment rather than a static state conferred by the platform. Gibbs, Ellison, and Lai \cite{ref22} investigated uncertainty reduction strategies in online dating, finding that personal security concerns, misrepresentation concerns, and recognition concerns all predicted the frequency of information-seeking behaviors. Corriero and Tong \cite{ref14} extended this work to mobile dating applications, showing how users of Grindr managed uncertainty through information seeking and concern monitoring. Hallam, De Backer, and Walrave \cite{ref26} examined the role of trust in the transition from online to offline dating, finding that the decision to meet in person is shaped less by accumulated trust than by a combination of practical and emotional factors. Coduto and Fox \cite{ref11} analyzed how affordances of mobile dating apps shape romantic relationship initiation and escalation, revealing paradoxical beliefs about trust and authenticity.

The concept of ``warranting''---the degree to which self-presented information can be verified through external sources---has been influential in understanding how online daters assess credibility \cite{ref18}. The warranting principle and the broader concern with credibility assessment are well established in prior HCI and CSCW research, which has long examined how people evaluate the reliability of online identities and information through cues that are costly to fake, cross-referencing across systems, and the affordances of the platforms that mediate self-presentation \cite{ref13,ref34}. Users seek warrants for others' claims through strategies such as requesting additional photos, conducting social media searches, and arranging video calls or in-person meetings. Platform affordances play a key role in shaping these verification practices: Comunello, Parisi, and Ieracitano \cite{ref13} showed how the design of dating apps mediates users' negotiation of social scripts and self-presentation norms.

Beyond individual strategies, research on collaborative sensemaking in online communities suggests that users collectively construct knowledge through shared interpretation of experiences. Mamykina, Nakikj, and Elhadad \cite{ref34} documented how members of online health forums engage in collective sensemaking to navigate uncertainty, a process that parallels the community-based risk recognition we observe among gay dating app users.

Our study builds on this literature by documenting how trust assessment operates in practice on gay dating platforms. We find that users develop layered, probabilistic strategies for evaluating others' credibility, combining textual consistency checks, photo verification, real-time interaction, and behavioral observation over time. Moreover, we show that these strategies are not purely individual inventions but are shaped by community knowledge: users learn from others' experiences, adopt shared heuristics, and collectively refine their understanding of what constitutes a trustworthy versus a suspicious interactional pattern.

\section{Method}

\subsection{Research Design}
This study adopted a qualitative research approach to understand gay male users' interaction experiences on dating applications and how they identified and responded to potential risks in platform-based interactions. Because such experiences are often embedded in specific interactional situations and participants' own interpretations, we chose semi-structured interviews as the primary method in order to obtain detailed accounts of participants' platform experiences. The interview protocol covered themes including platform use background, interaction experiences, risk judgment and coping strategies, and the circulation of experience within the community, aimed at revealing the judgments that users gradually developed through platform interaction and how these were shared and reinforced through community exchange.

\subsection{Participants and Recruitment}
This study recruited 22 male participants, ranging in age from 19 to 60. All had experience using gay social or dating applications, and most used multiple platforms simultaneously; those most frequently mentioned included Aloha, Blued, Fanka, and Soul. Length of use ranged from 1 to 9 years, with most participants having over three years of experience---sufficient for reflecting on how their understanding of interaction risks had evolved over time. Participants came from both urban and rural backgrounds with diverse occupations and educational levels, helping capture variation in how different users experienced platform-based interaction. Detailed participant information is provided in Appendix A.

Participants were recruited through community referral and snowball sampling. Initial contacts were made through personal networks; some participants then referred other eligible users. Because the research involved sensitive identities and private experiences, introductions through familiar social ties helped reduce concerns and increase willingness to participate. The only inclusion criterion was prior experience using gay dating or social applications. Although this criterion was deliberately broad, it was directly tied to our focus on deception: because deceptive encounters---ranging from misrepresented profiles to outright scams---are an ordinary by-product of sustained app use rather than a rare event, anyone who ``has used the apps'' over time is highly likely to have personally encountered or actively guarded against such situations. Relevant experience was therefore not assumed but verified during a brief pre-interview screening, in which we confirmed that each prospective participant had engaged in at least several rounds of actual interaction with other users (e.g., extended chatting, arranging or attending an offline meeting) rather than merely registering an account. This requirement, combined with the predominance of participants with more than three years of use, ensured that informants could speak from concrete first-hand experience with interactional risk. Potential participants were informed that the study focused on interaction experiences and associated risks, that no real names or identifying information would be required, and that interviews would proceed only after explicit consent was given. Throughout this paper we use \emph{deception} as an umbrella term for any intentional effort by another user to create a false impression---including profile misrepresentation, catfishing, romance or financial scams, and concealment of motives. Within this broader category, we reserve the term \emph{suspicion} for the participant's own anticipatory, often pre-verification state of mind: a felt uncertainty or wariness that a given interaction \emph{might} be deceptive, which precedes and may or may not be confirmed as actual deception. Suspicion thus denotes a perceiver-side orientation toward possible deceit, whereas other forms of deception denote the deceptive act itself; distinguishing the two lets us analyze how users manage risk even when no deception is ever proven.

\subsection{Ethical Considerations}
The study was approved by the ethics review board of an anonymized institution. Before each interview, participants were informed of the study's purpose, their right to refuse any question or withdraw at any time, and the ways data would be used. With consent, interviews were audio-recorded and subsequently anonymized. No real names, contact information, or identifying details appeared in the research records or this paper; participants are referred to by numerical identifiers throughout. Platform interaction scenarios mentioned during interviews were appropriately processed during transcription to prevent identification of either participants or third parties. Each participant received 50 RMB as compensation.

\subsection{Positionality and Researcher Reflexivity}
Following calls to make researcher positionality explicit in qualitative work---and especially in queer research and research situated in the Chinese context~\cite{ND-pillow2003,ND-kong2016,ref31}---we reflect on how our identities, commitments, and locations shaped this study. The research team is composed of members who variously identify as part of, or as allies to, LGBTQ+ communities, and who are themselves embedded in mainland Chinese social and digital environments. This insider familiarity gave us linguistic fluency in the community's evolving vocabulary, sensitivity to the coded and euphemistic registers participants use to navigate platform moderation, and access to recruitment networks that participants trusted. At the same time, we remained attentive to the ways our shared backgrounds could foreclose curiosity, leading us to treat partial perspectives as self-evident; we therefore practiced reflexive memoing and regular team debriefs to surface and interrogate our assumptions during data collection and analysis.

We further acknowledge the asymmetries within the team and between researchers and participants. Differences in gender identity, sexual orientation, regional origin, urban--rural background, and socioeconomic position meant that no single ``insider'' standpoint could speak for the diversity of queer experience in China, and we were mindful not to flatten participants' accounts into our own. Conducting queer research under conditions of platform censorship and uneven legal recognition also entailed a duty of care that exceeded procedural consent: we prioritized participants' safety in how we phrased questions, stored data, and reported findings, and we read our interpretations against participants' own framings rather than imposing external theoretical labels. We name these commitments not as a guarantee of objectivity but as an account of the situated, partial, and value-laden vantage from which this knowledge was produced~\cite{ND-haraway1988}.

\subsection{Data Collection}

This study used semi-structured interviews as the primary method of data collection \cite{ref1,ref15,ref29}. Interviews were conducted primarily through online voice or video calls to facilitate participation across different cities; each lasted approximately 50--60 minutes and was audio-recorded with participants' consent before transcription and anonymization.

The interviews proceeded from less sensitive to more sensitive topics, moving through four stages: platform use background, problematic interaction experiences, risk judgment and coping strategies, and community experience and learning. Within each stage, researchers asked participants to describe specific situations, identify what triggered their suspicions, explain how they evaluated credibility and verified identity, and reflect on how experiences circulated among community members. The full interview protocol is provided in Appendix B.

\subsection{Data Analysis}
This study employed thematic analysis \cite{ref5,ref6}. Transcripts were independently coded by two researchers, who assigned descriptive labels to segments relevant to the research questions---covering problematic interaction situations, judgment cues, coping behaviors, and community experience. Codes were then compared, merged, and iteratively refined through discussion until thematic saturation was reached. The resulting higher-level analytical structure informed the three main findings themes: types of deceptive interactions, users' risk recognition and coping strategies, and community-based mechanisms for generating and circulating judgment knowledge. Representative interview excerpts are used throughout to demonstrate how each theme is grounded in participants' accounts.

\section{Findings}
Our findings are organized around three interconnected dimensions: the interactional structures that render deception possible and difficult to detect (Section 4.1), the judgment practices users develop in response to persistent uncertainty (Section 4.2), and the community-based processes through which these judgment capacities are collectively generated, circulated, and sustained (Section 4.3). Together, these three dimensions trace a path from interactional structure to individual practice to collective knowledge production, revealing that navigating deception on gay dating applications is neither a matter of detecting isolated falsehoods nor a purely individual skill, but a socially embedded and continuously evolving process.

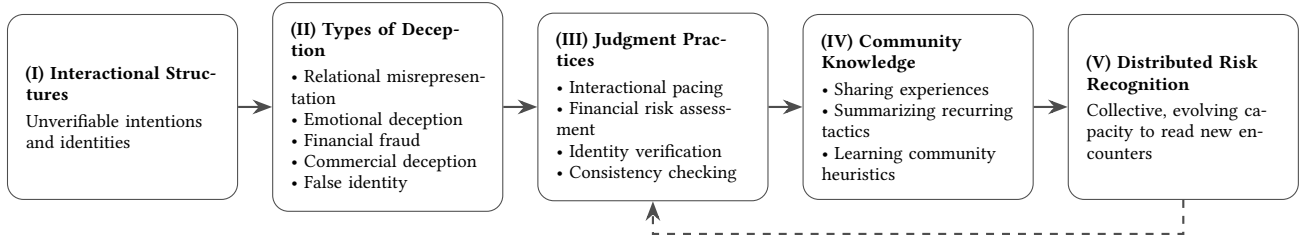
\begin{figure}[t]
\centering
\begin{tikzpicture}[
  box/.style={draw=black!70, rounded corners=6pt, align=left, font=\footnotesize, inner sep=7pt, text width=2.55cm, minimum height=2.45cm},
  head/.style={font=\footnotesize\bfseries},
  arr/.style={-{Stealth[length=2.6mm]}, thick, black!70},
  node distance=0.45cm]
\node[box] (s1) {\textbf{(I) Interactional Structures}\\[2pt]Unverifiable intentions and identities};
\node[box, right=of s1] (s2) {\textbf{(II) Types of Deception}\\[2pt]$\bullet$ Relational misrepresentation\\$\bullet$ Emotional deception\\$\bullet$ Financial fraud\\$\bullet$ Commercial deception\\$\bullet$ False identity};
\node[box, right=of s2] (s3) {\textbf{(III) Judgment Practices}\\[2pt]$\bullet$ Interactional pacing\\$\bullet$ Financial risk assessment\\$\bullet$ Identity verification\\$\bullet$ Consistency checking};
\node[box, right=of s3] (s4) {\textbf{(IV) Community Knowledge}\\[2pt]$\bullet$ Sharing experiences\\$\bullet$ Summarizing recurring tactics\\$\bullet$ Learning community heuristics};
\node[box, right=of s4] (s5) {\textbf{(V) Distributed Risk Recognition}\\[2pt]Collective, evolving capacity to read new encounters};
\draw[arr] (s1) -- (s2);
\draw[arr] (s2) -- (s3);
\draw[arr] (s3) -- (s4);
\draw[arr] (s4) -- (s5);
\draw[arr, dashed] (s5.south) -- ++(0,-0.45) -| (s3.south);
\end{tikzpicture}
\caption{Analytical framework: interactional structures render intent and means unverifiable (Section~4.1); users respond with individual judgment practices (Section~4.2); these are pooled into community knowledge (Section~4.3), yielding a distributed risk-recognition capacity that feeds back into individual judgment (dashed arrow).}
\label{fig:nd-framework}
\end{figure}

Figure~\ref{fig:nd-framework} visualizes this analytical framework: platform-level interactional structures make intentions and identities unverifiable; within these structures users encounter distinct types of deception; individuals respond with judgment practices; these practices are pooled into community knowledge; and the result is a distributed capacity for risk recognition that feeds back into how new encounters are judged.

\subsection{Interactional Structures That Render Intent and Means Unverifiable: A Means- and Intent-Based Typology of Deception}

We organize the following analysis around a two-dimensional typology, and we begin by defining its two axes, since every category below is located on both. The first axis concerns the \emph{means} of deception: the interactional resources through which a false impression is produced and sustained---accelerating the pace of intimacy, deliberately maintaining ambiguity, framing the relationship in a particular register, or constructing a fabricated identity. The second axis concerns the \emph{intent} of deception: the concealed goal that the deceptive interaction ultimately serves---sexual access, emotional occupation, financial extraction, or commercial promotion. The two axes are analytically distinct (orthogonal) in the sense that they vary independently rather than being opposite ends of one continuum: every concrete deception involves \emph{both} a means and an intent; the same means can serve different intents (an accelerated-intimacy script may precede a sexual, financial, or commercial goal); and the same intent can be pursued through different means. Each type we present is therefore not a point on a single axis but a recurring, culturally recognizable \emph{pairing} of a characteristic means with a characteristic intent. The distinction also does empirical work: participants' moral evaluations of deception tracked the intent axis (what the other party was ``really after''), whereas their practical detection and verification strategies necessarily operated on the means axis (the observable pacing, framing, and self-presentation of the interaction), because intent itself is never directly observable. A central condition shaping participants' experiences was that the other party's true intentions and the authenticity of their self-presentation could rarely be directly verified within the interaction itself. The deceptions they described were not objectively false statements checkable against external evidence but were embedded in the relational dynamics of interaction: in the pacing of emotional escalation, the alignment between expressed and actual intent, and the strategic use of relational scripts to achieve concealed goals. 

Table~\ref{tab:nd-deception-typology} summarizes the four means--intent pairings and the verification strategies participants attempted against each; identity construction (Section~4.1.5) appears as a cross-cutting means that can be coupled with any of the four intents.

\begin{table}[t]
\centering
\caption{Deception types as recurring pairings of a characteristic means with a concealed intent (Sections~4.1.1--4.1.5), together with the interactional cues participants learned to read and the verification practices they attempted. Identity construction is a cross-cutting means that can support any of the four intents.}
\label{tab:nd-deception-typology}
\footnotesize
\begin{tabular}{p{0.155\linewidth}p{0.115\linewidth}p{0.135\linewidth}p{0.24\linewidth}p{0.24\linewidth}}
\toprule
\textbf{Deception type} & \textbf{Characteristic means} & \textbf{Concealed intent} & \textbf{Recognizing interactional cues} & \textbf{Verification practices} \\
\midrule
Relational misrepresentation for sexual access (4.1.1) & Accelerated intimacy scripts & Sexual access (\emph{piao-pao}) & Expressing intimacy rapidly; avoiding discussion of long-term intentions; withdrawing after achieving sexual goals & Slowing interaction; delaying offline meetings; observing consistency of relational intentions \\
\addlinespace
Sustained emotional deception (4.1.2) & Prolonged ambiguity; deferral of definition & Emotional occupation without commitment & Escalating emotional commitment; providing inconsistent personal narratives; making contradictory promises & Extending interaction; cross-checking narratives; evaluating long-term consistency \\
\addlinespace
Financial fraud (4.1.3) & Investment and hardship narratives within a relational frame & Financial extraction & Introducing investment opportunities; requesting financial assistance; shifting conversations toward money & Refusing financial transfers; verifying external information; treating financial requests as warning signals \\
\addlinespace
Commercial deception disguised as ordinary users (4.1.4) & Ordinary-user posture; everyday sociable chat & Commercial promotion or redirection & Redirecting conversations toward products or services; promoting commercial content; sharing external links & Examining account history; comparing interaction patterns; disengaging from suspicious accounts \\
\addlinespace
False identity construction (4.1.5) & Fabricated profile and persona & Cross-cutting: serves any of the above & Presenting inconsistent profile information; using misleading photographs; providing contradictory self-descriptions & Requesting additional photographs; conducting cross-platform searches; arranging video calls or offline verification \\
\bottomrule
\end{tabular}
\end{table}%

\subsubsection{Accelerating Intimacy to Achieve Sexual Access}

This was the most common form of deception mentioned in the interviews, and all participants referred to it. Participants generally called it pian pao (``deceiving someone into sex''; in Chinese, \emph{pian pao}, written `\zh{骗炮}', literally ``swindling a shot,'' a colloquial term for inducing a sexual encounter under false pretenses), referring to the use of different forms of verbal deception on dating apps to induce the other person to have sex. In the typology of documented deceptive practices, pian pao corresponds to a form of \emph{sexual-access} or ``body-fishing'' deception, in which fabricated relational sincerity is instrumentalized to obtain physical intimacy~\cite{ref45}. It usually centers on misleading relational intent: for example, the other person presents themselves as being interested in a long-term relationship, stable dating, or serious emotional connection, while their actual goal is simply to have sex as quickly as possible. 

Sixteen of the 22 participants noted that gay dating platforms such as Blued and Aloha do not themselves exclude short-term sexual encounters, and that many users already know that some people join these platforms specifically for short-term or purely sexual relationships. 

\begin{quote}
P2: ``If someone makes it clear from the beginning that it is just about sex, that is actually okay. People can decide for themselves whether they accept that. The problem is that some people will say they want a serious emotional relationship, talk a lot about the future, talk a lot about life, as if they really want to develop something, but in the end it is just to get you into bed more quickly.''
\end{quote}

In participants' view, the key to this kind of deception was not simply ``telling lies,'' but the way it relied on switching between two different relational scripts available on the platform. 

One participant noted that the other person would often intensify intimacy quickly before sufficient mutual understanding had been established:

\begin{quote}
P6: ``Some people will immediately say things like, `You're different from other people,' or `Talking with you feels really comfortable,' and make it seem like they genuinely want to get to know you seriously. Looking back now, that kind of very rapid closeness can actually be a signal.''
\end{quote}

Other participants pointed out that one reason pian pao is difficult to clearly name as ``deception'' is that it can still be interpreted on the surface as ``the relationship didn't work out'' or ``they later lost interest,'' which makes it hard for those affected to obtain a clear judgment from the community:

\begin{quote}
P11: ``The hardest part is that it is difficult to say for sure whether he was deceiving you, because it is not as explicit as asking you for money. But you feel that all that seriousness and care before was serving the same purpose.''
\end{quote}

Taken together, participants' accounts suggest that pian pao was experienced not as a mere hookup, but as a sexual-access deception that achieves short-term sexual goals by borrowing the symbols, pace, and emotional expressions of long-term relationships.

\subsubsection{Sustaining Ambiguity to Maintain Emotional Engagement}
Different from relational misrepresentation for sexual access, another risk frequently mentioned by participants was a more sustained and more ambiguous form of emotional deception, raised by 15 participants. In this kind of interaction, the other person uses long-term chatting, steady companionship, emotional responsiveness, and ambiguous expressions of intimacy to gradually make the other person believe that the interaction may develop into a real relationship. 

One participant said:
\begin{quote}
P4: ``He would come talk to me every day. He never missed saying good morning or good night. He would ask me how I was doing and also comfort me. After a while, you naturally feel like the relationship is moving forward. But every time I brought up meeting or defining the relationship, he would dodge it.''
\end{quote}

Unlike pian pao (\textit{piao-pao}, \zh{骗炮})---a colloquial term for soliciting an in-person encounter under romantic pretenses purely to obtain sexual access, a form of `body-fishing' deception in which the deceiver's goal is a short-term physical outcome rather than the relationship implied, emotional deception may not end quickly but persists through repeated cycles of drawing close and withdrawing. 

One participant remarked:
\begin{quote}
P9: ``Sometimes you know that this person may not really want to develop anything, but because he keeps being there, it is hard to completely pull yourself out. That is the most exhausting part.''
\end{quote}

Other participants noted that this kind of emotional deception is not always driven by clearly malicious intent---some people may simply want companionship or validation without taking on relational responsibilities. Precisely for this reason, it is harder to name within the community; many participants described such people not as ``scammers'' but as ``keeping you hanging,'' ``fishing,'' or ``just enjoying being liked.''

For example:
\begin{quote}
P16: ``You cannot necessarily say he is a liar, because he may not be consciously designing it that way. But he is definitely using your expectations to keep you invested.''
\end{quote}

\subsubsection{Leveraging Relational Framing for Resource Extraction}
In participants' accounts, the form of deception most readily named as ``fraud'' was interaction oriented toward financial gain, although only seven participants mentioned it. Unlike crude scams in which someone immediately asks to borrow money, the financial fraud participants described was typically processual.  

One participant recalled:
\begin{quote}
P11: ``At the beginning he would not talk about money at all. He would just come chat with you every day, ask what you were doing, whether you had gone to sleep at night, and it all felt pretty normal. After chatting for about a week or two, he started saying that he had run into some difficulties recently and asked whether I could help.''
\end{quote}

This kind of fraud often relied on several narrative strategies to achieve financial extraction. The first was a ``short-term difficulty'' narrative, such as a broken phone, overdue rent, problems at work, or illness in the family.  The second was an ``investment or money-making opportunity'' narrative, in which the person displays apparently credible work experience, investment knowledge, or resources in order to induce the other party to participate in a promised return. This investment (\begin{CJK}{UTF8}{gbsn}投资\end{CJK}) framing maps directly onto the documented ``pig-butchering'' (\begin{CJK}{UTF8}{gbsn}杀猪盘\end{CJK}) romance--investment scam pattern, in which a cultivated intimate bond is leveraged to channel the victim toward a fraudulent trading or investment platform~\cite{ND-Wang2023,ND-FBI2022,ND-aissyach2025}. The third was a ``mutual support within a relationship'' narrative, which frames financial assistance as a natural form of support within what is presented as an already intimate bond.
\begin{quote}
P16: ``He said he was on a business trip in another city and that his bank card had transfer limits, so he hoped I could cover it for him first. At first I really hesitated, because we had been chatting well and it felt like maybe I was just helping someone out.''
\end{quote}

Some participants also noted that the other person would not necessarily directly ask to borrow money, but would instead repeatedly hint at hardship in the hope that the participant would volunteer help:
\begin{quote}
P17: ``He would not directly say, `Transfer money to me.' He would just keep talking about how miserable and how difficult things were for him, as if he wanted you to soften on your own.''
\end{quote}

Participants came to recognize that scammers exploited not only information asymmetry but also users' expectations of intimate relational scripts: financial fraud was effective because that story was embedded within what appeared to be a relationship in formation, reactivating trust, sympathy, and felt responsibility.

\subsubsection{Using Interaction as Cover for Commercial Redirection}
Participants also frequently mentioned another category of accounts that seemed to be ``not social interaction, but business'': on the surface, these accounts disguised themselves as ordinary users and engaged in apparently natural conversation and interaction, but their actual purpose was to drive traffic, sell products, promote services, recruit people, or use the platform as an entry point into some form of commercial activity. Participants commonly described such behavior as ``hanging up a sheep's head while selling dog meat'' (`\zh{挂羊头卖狗肉}', a Chinese idiom denoting the practice of advertising one thing while actually delivering another). This framing maps onto forms documented in the online romance-fraud literature, in which a benign relational presentation is deliberately constructed to mask an underlying commercial or extractive motive~\cite{ref49}.

One participant noted:
\begin{quote}
P22: ``At first they look no different from ordinary people and will also chat about daily life. But after a while they start saying whether you want to add another platform, or recommending that you go check something out.''
\end{quote}

Unlike financial fraud, marketing accounts batch-process opportunities for contact, converting a space meant for relationships into an entry point for commercial resources.  
\begin{quote}
P14: ``The most annoying thing is not that he is doing marketing. It is that he is clearly not here to meet people, but pretends as if he is seriously chatting with you.''
\end{quote}

Marketing behavior thus constituted a risk not because of commerce itself, but because it entered relational spaces under false pretenses, a form of disguised commercial activity that blurred the boundary between ``meeting people'' and ``being redirected.''

\subsubsection{Identity Construction as a Supporting Mechanism}
While the preceding sections examined forms of deception differentiated by their underlying intent—sexual access (i.e., piao-pao / `\zh{骗炮}', sexual-access or `body-fishing' deception, in which a fabricated persona is used to obtain a sexual encounter), emotional occupation, financial extraction, and commercial redirection (i.e., the `\zh{挂羊头卖狗肉}' pattern, literally `hanging a sheep's head while selling dog meat,' here a form of disguised commercial activity in which dating intent is feigned to channel the other person toward a paid product or service)—false identity operates at a different level: it is primarily a means-based form of deception rather than an intent-based one.  Whereas the previous categories were distinguished by purpose and outcome, participants understood false identity as a means of carrying out deception; in the vocabulary of our typology (Section~4.1), it occupies the means axis and can be coupled with any of the four intents documented above.  This was likewise described as a very common form of deception and was mentioned by all participants.

What participants referred to as ``false identity'' encompassed multiple layers: using photos differing substantially from one's actual appearance; fabricating age, height, body type, sexual role, occupation, educational background, income level, and emotional status. 

One participant said:
\begin{quote}
P2: ``Some people package themselves really well, as if they have a great job, a very high-quality lifestyle, and very polished photos. But once you talk more with them, you find that a lot of it does not hold up under questioning.''
\end{quote}

Another participant pointed out that false identity is not always a complete impersonation of another person, but can also involve a strategic reconstruction of the self:
\begin{quote}
P15: ``It is not that he is completely not that person. It is that the version he gives you has been put together specifically to attract you.''
\end{quote}

At the level of photos, participants most frequently mentioned excessive photo editing, the use of photos from many years earlier, deliberately selected angles that do not reflect one's actual condition, and even directly stealing another person's profile picture.  

One participant recalled:
\begin{quote}
P11: ``After meeting in person, the difference was huge. It is not that the person was necessarily unattractive, but you realize that the image he gave you was simply not what he looked like in real life.''
\end{quote}

With regard to profession and social status, participants also noted that some people deliberately used labels such as high income, high education, government or institutional employment, or overseas study background to establish credibility and attractiveness.  

In addition, some participants mentioned deception regarding sexual roles.
\begin{quote}
P22: ``He wrote on his profile that he was 0.5 or 1 [a role classification commonly used in the Blued community among gay men, mainly including 0, 1, and 0.5, with different numbers often indicating the role one is willing to play in intimate and sexual relationships], but when we actually met he was completely 0.''
\end{quote}

Participants did not regard all forms of self-enhancement as serious deception; choosing flattering photos was seen as nearly normal. But when packaging moved beyond ``showing one's best side'' and clearly misled the other person's judgment, false identity ceased to be self-presentation and became a tool of deception. Because it often serves sexual deception (piao-pao), emotional deception, or financial fraud, it has a strongly ``prior'' character in participants' experiences: many later risks begin with a carefully constructed false self.

\subsection{Users' Coping Practices Under Interactional Uncertainty}

Precisely because others' intentions and self-presentation cannot be directly verified, users are compelled to develop their own strategies for navigating persistent uncertainty. Through sustained platform use, participants developed verification and coping practices embedded in everyday interaction and mobilized across the trajectory from ``starting to chat'' to ``considering meeting in person.''

These strategies were not designed to obtain definitive certainty. Rather, participants engaged in what can be characterized as probabilistic inference: continuously integrating multiple signals---textual consistency, photo authenticity, interactional pace, behavioral responses to sensitive topics---into an evolving assessment of whether the interaction was ``sufficiently credible'' to proceed. Crucially, these practices map onto the means/intent typology developed above: ``verifying identity'' targets the \emph{means} of deception (false self-presentation), whereas ``judging intent'' and ``recognizing commercial purpose'' target the \emph{intent} behind it (emotional occupation, financial extraction, or commercial redirection). Participants combined both forms of work, treating identity cues as a first-order filter and intent cues as the higher-stakes judgment on which decisions to continue or meet ultimately rested.

\subsubsection{Judging and Testing Relational Intent}
Judging relational intent is often more difficult because it does not exist as an ``objective fact'' that can be directly verified.  For this reason, when facing risks of sexual deception and emotional deception, participants relied more on a combined reading of interactional pace, intensity of emotional expression, conversational topics, and consistency of behavior.  Whereas the previous strategies targeted the \emph{means} of deception, the cues described here address its \emph{intent}, the analytic distinction we carry through this section.

Fifteen participants mentioned that they were especially wary of ``closeness that develops too quickly.'' When the other person frequently used strongly affirmative emotional language shortly after getting to know them—for example, ``you're different from other people,'' ``I rarely meet someone like you,'' or ``I think we're a great match''—participants would not simply interpret this as romance, but rather as a signal requiring further observation.  In short, emotional escalation outpacing mutual understanding read as a marker of intent rather than sincerity.

One participant said:
\begin{quote}
P1: ``Now, if someone enters that very ambiguous, very serious, very emotionally intense state too quickly, I instead remind myself to slow down. Because you actually do not know each other at all yet.''
\end{quote}

Accordingly, when judging relational intent, participants often paid attention to whether ``verbal promises'' and ``real-world progression'' were consistent.  Sexual-deception risk surfaced as conversation steered prematurely toward bodies or meeting places, while emotional-deception risk surfaced as sustained ambiguity and avoidance of relationship confirmation.

For example, one participant said:
\begin{quote}
P14: ``If he keeps saying he wants something serious, but every time the conversation ends up turning toward sex, then I basically know where his focus is.''
\end{quote}

Another participant emphasized that delay and vagueness are themselves signals:
\begin{quote}
P16: ``Some people will keep chatting with you in a very intimate way, but every time you bring up meeting or seriously moving things forward, he drags it out. After enough time, you know that this kind of person probably does not really want to develop anything.''
\end{quote}

  By slowing the pace or withholding intimacy and then watching for sudden disengagement, they probed intent directly. 

One participant said:
\begin{quote}
P21: ``Later I would deliberately slow things down a bit, like not meeting so quickly. If the other person cools off right away, that actually says a lot.''
\end{quote}

  This experiential, intent-oriented judgment thus complemented the means-oriented verification discussed earlier.

\subsubsection{Identifying and Refusing Financial Risk}
Compared with relational intent, financial risk can in some sense produce a clearer boundary, because ``talking about money'' is often a relatively obvious turning point.   In terms of the means/intent typology, an early monetary request functioned as a salient \emph{means}-level cue from which participants inferred extractive intent, making this the most legible risk signal in their accounts.

One participant said:
\begin{quote}
P16: ``Now for me, as long as we haven't met many times or the relationship is nowhere near that level, once money enters the conversation, it's basically over.''
\end{quote}

  Here participants decoupled inferred intent from relational means, refusing to let conversational closeness be read as real-world obligation.

In addition, some participants mentioned several common patterns they attended to when identifying financial risk, such as a rapid shift into narratives of hardship, repeated emphasis that the other person ``would never normally ask for help,'' attempts to manufacture urgency, unwillingness to provide verifiable real-world information, or the use of emotional pressure after a monetary request was refused.  These cues were read together as a \emph{means}-level signature of extractive intent: the issue was less the request itself than its packaging as morally hard to refuse.

\subsubsection{Recognizing Commercial Promotion Accounts}
Among all forms of deception, participants generally described this as the easiest to identify and cope with. Within the means/intent typology, marketing accounts were distinctive in that detection hinged on intent rather than means: the point of judgment was not whether the other party was a ``real human,'' but whether that account had come to the platform to establish a normal relationship.  Once participants detected a clear tendency toward traffic redirection, product selling, group recruitment, promotion of paid content, or a persistent effort to steer the conversation toward external commercial resources, they would usually classify the account as a marketing account and terminate the interaction.

One participant said:
\begin{quote}
P19: ``Some people try to get you onto another platform really quickly, saying it is inconvenient to chat here. Now, the moment I hear that, I become more alert.''
\end{quote}

Some participants also summarized a more general criterion: ordinary conversations aimed at getting to know someone usually revolve around each other, whereas marketing-oriented conversations quickly shift the focus from ``who you are'' to ``where to go,'' ``what to look at,'' ``what to join,'' or ``what to buy.''  The diagnostic cue, then, was directional intent rather than any single utterance.

Some participants stated that once they detected this directional pattern, they ended the conversation outright rather than testing further.

\subsubsection{Verifying the Authenticity of Identity}
Faced with the risk of identity deception, participants usually verified the authenticity of the other person's identity in multiple gradual ways.   Read through the means/intent typology, these strategies progressively raised the cost of sustaining a fabricated identity, targeting deceptive \emph{means} while inferring underlying \emph{intent}.

First, fifteen participants said that they checked the other person's basic identity information by asking follow-up questions about details during chat.   Such questions functioned as a low-cost, non-confrontational test of credibility, judged by the coherence, detail, and responsiveness of the other person's account.

Another participant mentioned:
\begin{quote}
P12: ``Some people write a great profile, but once you start chatting, you realize a lot of it is empty. If you ask a slightly more specific question, they start becoming vague or their story becomes inconsistent. For example, I once met a user who wrote on his profile that he liked European philosophy. Then I asked him what Critique of Pure Reason was, and he had never heard of it. So I felt that his claim about liking philosophy was not very credible.''
\end{quote}

  Verification thus aimed less at catching a single lie than at testing whether a narrative sustained the continuity of ``real life.''

In addition to checking textual information, photos were also an important basis on which participants judged identity authenticity.   

A common strategy was to ask for more photos, especially everyday photos taken in different settings or at different times.  Consistency across photos was itself a clue, with mismatched features or a single standardized selfie template raising suspicion.

One participant said:
\begin{quote}
P21: ``If there are only one or two especially attractive photos, I usually feel that is not enough. I want to see what he looks like in everyday photos, or in different states.''
\end{quote}

On this basis, some participants also used real-time photos for further confirmation.  Asking for a ``photo taken right now,'' sometimes with a specified action or setting, did not eliminate risk but confirmed that the person in the photo was the one currently chatting.

For example:
\begin{quote}
P6: ``If we are getting close to meeting up, I sometimes ask him to send a photo taken right now. For example, send one at the office or at home. That makes me feel much more reassured.''
\end{quote}

In addition, some participants used video calls for further verification.  As a stronger form of confirmation, video simultaneously verified appearance, real-time presence, and interactional continuity, but had to be proposed only after some relational foundation existed, since an early request could read as distrust.

One participant explained:
\begin{quote}
P13: ``Video is the most direct way, but it is also not something you can bring up right away. Usually it is after we have chatted for a while and I feel like we might meet that I will ask whether we should video first.''
\end{quote}

As for other kinds of information---such as educational background, income, emotional status, and relationship history---participants generally said that they did not have particularly effective ways of verifying them, and instead relied on prolonged observation and inference during ongoing conversation.
\begin{quote}
P4: ``Education and income---there is no real way to verify those. I cannot ask him to send me his pay slip, and even then I would not know whether it was really his. But I believe that if it is fake, sooner or later it will show. If you keep talking long enough, you will eventually notice places where the logic breaks down.''
\end{quote}

Overall, participants adopted a gradually escalating verification strategy: beginning with checking details in conversation, moving to observing photo consistency, then confirming through real-time photos or video.  These practices were social as much as technical, balancing risk reduction against a workable interactional atmosphere, and they illustrate how users read deceptive \emph{means} to infer \emph{intent} across escalating verification thresholds.

\subsection{The Community-Based Generation of Risk Recognition Capacities}

The judgment practices described above---graduated, probabilistic, and signal-based---raise a further question: where do these capacities come from?  They are progressively generated, circulated, and stabilized through community-based processes of experience sharing, collective interpretation, and pattern abstraction.  In effect, the community operates as a decentralized mechanism for producing and updating the practical knowledge users draw upon, much of which is organized around reading others' \emph{means} and inferring their \emph{intent}.

Nineteen participants stated that they had not learned these judgments on their own.   Participants saw formal mechanisms (reporting, banning, real-name verification) as ill-suited to the gray-area cases---pian pao, prolonged emotional ambiguity, or disguised marketing---where harm hinges on contested readings of means and intent rather than on bright-line rules; community exchange filled this gap.

\subsubsection{Circulation and Collective Interpretation of Interactional Experiences}
Many participants noted that when they themselves, or their friends, encountered suspicious accounts, marketing behavior, experiences of financial fraud, or even obvious sexual or emotional deception, their first reaction was often not to report the account immediately to the platform, but rather to talk about it first with acquaintances in the community, in WeChat groups, or in chats with friends.  Such circulation served varied intents---warning others, seeking confirmation, or obtaining comfort.  Regardless of intent, sharing moved an individual experience into a broader interpretive network, where the means and intent of suspected deception could be jointly assessed.

One participant said:
\begin{quote}
P5: ``If I run into something especially outrageous, I usually tell my friends first and see whether they have encountered anything similar.''
\end{quote}

In this circulation process, screenshots, profile photos, chat histories, and claimed information all became materials that could be analyzed collectively.  Sharing thus functioned as collaborative verification rather than mere venting, with peers reading these materials to infer deceptive means and intent.

For example:
\begin{quote}
P19: ``Sometimes when you are inside it yourself, you cannot see clearly. But a friend looks at it and says this really looks like one of those taolu [routine patterns], and I suddenly realize what is going on.''
\end{quote}

  Where forms such as pian pao or emotional deception resist formal establishment under platform rules, community discourse can still name the felt experience of being misled, partly compensating for the interpretive gap left by formal mechanisms.

\subsubsection{Formation of Shared Judgment Heuristics (Taolu)}
As individual experiences were repeatedly narrated and compared, participants gradually summarized them as taolu (routines/patterns). Community discussion did not merely repeat individual cases but extracted commonalities across opening lines, emotional-escalation rhythms, and profile cues. In effect, these taolu encode the means/intent typology: each pattern bundles a recognizable interactional means with an inferred underlying intent.

One participant said:
\begin{quote}
P22: ``After people in the community talk enough, you realize that many things actually follow fixed patterns. It is not just something that happened to you alone.''
\end{quote}

 For example, the community might summarize ``rapidly building closeness---creating a sense of uniqueness---pushing quickly toward meeting'' as a common path of pian pao; ``steady companionship---avoiding meeting---keeping someone hanging for a long time'' as a typical rhythm of emotional deception; ``apparently ordinary chat---quickly redirecting to external platforms'' as a marketing account's mode of operation; and ``building familiarity---narrating hardship---creating urgency'' as a common script of financial fraud. Each script thus pairs a characteristic means with a distinct intent, mapping the typology onto concrete sequences.

\subsubsection{Stabilization and Codification of Community Rules}
On the basis of the circulation of experience and the summarization of patterns, 15 participants further noted that the community gradually formed some rules of judgment that may not be formally written down, but were nonetheless widely followed.  Examples included: ``do not have financial exchanges before meeting in person,'' ``video before meeting if possible,'' ``be cautious when someone enters a passionate romantic mode too quickly,'' ``long-term ambiguous relationships that never move offline should be treated carefully,'' and ``accounts that quickly redirect you to outside platforms probably have another agenda.'' Read through the means/intent typology, these rules operate as heuristics that flag suspect \emph{means} (premature financial requests, off-platform redirection) and infer adversarial \emph{intent} from their co-occurrence.

One participant said:
\begin{quote}
P3: ``A lot of things are not taught to you explicitly by one person. It is just that everyone says them, and after hearing them enough, you realize they are default rules.''
\end{quote}

  Without guaranteeing safety, they supply a shared framework for action under uncertainty and recalibrate the community's normative boundaries around pacing, requests, and acceptable ambiguity.

 Some participants felt certain rules were too conservative---``not everyone who refuses to video is necessarily deceptive''---and others cautioned that heavy reliance on rules might obscure individual variation.

\subsubsection{Intergenerational Transmission and the Reproduction of Risk Knowledge}
Many participants reflected that they did not initially possess so many judgment resources.  Long-term immersion in community experience was essential, and the ability to recognize risk was learned through it. Crucially, what circulated was not a generic alertness but a shared typology for reading others' means and intent.

\begin{quote}
P17: ``A lot of my avoidance knowledge came from friends. For example, what kinds of people may just want to deceive you into sex, and what kinds of scripts marketing accounts often use---these are not things the platform teaches you. If it were not for my close friends, I might have had to be deceived by every one of these methods before I learned.''
\end{quote}

P17's account shows the community transmitting exactly the means/intent distinctions of our typology: which intents to suspect and which scripted means signal them. 

Risk learning is thus a cycle rather than one-way input: users enter through others' experience and later recirculate it, making intent-reading a collectively maintained practice.

\section{Discussion}

Taken together, our findings cohere into a single argument: the typology of practices we describe, the trust and verification work our participants perform, and the community knowledge they circulate are not three separate phenomena but three faces of one ongoing accomplishment. Participants do not merely classify resources, assess their credibility, and share what they learn as discrete steps; rather, the typology supplies the categories through which trust is negotiated, verification practices are themselves a form of distributed community knowledge, and the knowledge that circulates in turn reshapes which categories feel trustworthy in the first place. Read this way, what looks like individual sense-making is better understood as a collective infrastructure that participants continuously build, repair, and hand off to one another. This reframing matters because it relocates the design problem away from helping a lone user judge a single artifact and toward supporting the social process by which communities decide, together, what counts as reliable.

We position this argument in dialogue with queer CSCW and its critique of the figure of ``the user.'' Bardzell and Bardzell's call to reconfigure the user~\cite{ND-BardzellBardzell2015} reminds us that the verification practices we observed are enacted by situated, marginalized actors whose needs are poorly served by systems that presume a neutral, individuated information seeker; our participants' work is precisely the labor of making infrastructures legible and safe for people the default design imagination overlooks. Dourish's caution that ethnographic accounts resist translation into tidy ``implications for design''~\cite{ND-Dourish2006} tempers any move to read our typology as a checklist of features; instead it asks us to treat the community knowledge we documented as an analytic resource that reframes what design is for. Finally, Wong-Villacres and colleagues' attention to intersectional and assets-based perspectives~\cite{ND-WongVillacres2020} foregrounds how the trust work in our setting draws on, rather than compensates for, the existing strengths of these communities. Holding these three commitments together, we argue that supporting verification means supporting the queer, collective, and asset-rich practices through which trust is already being produced, not supplanting them with tools built for an idealized solitary user.

\subsection{Rethinking Deception: From False Information to Misaligned Interactional Logics}

This subsection consolidates former Sections 5.3 and 5.5 into a single argument about what our typology contributes to how CSCW conceptualizes deception. We first question the prevailing equation of deception with false information, and then show that the forms we documented are best differentiated not by their surface content but by the interactional logics through which they unfold.

Our findings also point to the need to broaden how deception is conceptualized in platform-mediated interaction. Most existing literature focuses on discrepancies between presented information and objective reality---inaccurate profiles or fabricated identities \cite{ref19,ref43,ref44}. But participants' experiences suggest that deception often lies not in false information per se, but in the misalignment between expressed and actual intentions. Presenting oneself as seeking a serious relationship while pursuing short-term sexual encounters, or maintaining prolonged emotional engagement without intention to develop the relationship, may involve no explicitly false statements yet can still be deeply experienced as deceptive. This shifts attention from factual accuracy to relational alignment \cite{ref24}: users navigate not only what others say about themselves, but what those others intend to do within the interaction. This intentional misalignment can be read through the lens of interpersonal deception theory, which treats deception as a strategic, interactive accomplishment rather than a discrete false claim \cite{ND-buller1996}, and it resonates with accounts of selective self-presentation in mediated contexts, where deception emerges from what is strategically withheld or implied rather than stated outright \cite{ref47}. Integrating these concepts, we frame intentional misalignment as a form of relational deception that conventional veracity-based definitions fail to capture.

The forms of deception identified in our data should not be treated as merely distinct categories, but as reflecting different underlying interactional logics. Relational misrepresentation toward sexual access relies on accelerating emotional intimacy to transition quickly into offline encounters. Emotional deception operates through prolonged ambiguity—sustained contact without progression or commitment. Financial fraud involves a sequential shift from ordinary interaction to monetary requests, leveraging relational context to legitimize them \cite{ref50,ref51}. Marketing-oriented accounts enter the relational space as ordinary users before redirecting toward commercial ends. Each form involves distinct ways of organizing interaction and managing expectations, which is why no single verification strategy can address all types of risk and why users must adopt differentiated responses. Conceptually, these logics can be integrated through the lens of \emph{relational framing}: each form of deception exploits the interactional expectations that a particular frame makes salient, such that the apparent diversity of tactics resolves into a common mechanism of manipulating how interaction is framed rather than the content exchanged within it \cite{ND-goffman1974}. Reading our findings through this frame clarifies why differentiated responses are not ad hoc reactions but principled adjustments to the specific framing each logic invokes.

\subsection{Making Do Under Uncertainty: Trust Work, Processual Judgment, and Community-Based Compensation}

This subsection consolidates former Sections 5.1, 5.2, and 5.4 into an account of how participants ``make do'' within structures they cannot change. Rather than achieving verification, they manage uncertainty continuously; rather than detecting discrete violations, they read unfolding trajectories; and where individual judgment reaches its limits, community knowledge compensates. The three practices are successive layers of a single adaptive repertoire.

While prior research often frames trust as a matter of verifying the truthfulness of others' self-presentation \cite{ref18,ref27}, our findings suggest that, in the context of gay dating applications, definitive verification is rarely achievable. Participants instead described continuously assessing, revising, and updating their judgments in response to unfolding interactional cues. Trust emerges not from the confirmation of objective facts but from a situated and provisional evaluation of risk under persistent uncertainty---a level of confidence ``sufficient'' to proceed to the next step. This reframes trust from a binary outcome into an ongoing process of uncertainty management \cite{ref14,ref22}, where judgments remain open to revision and are shaped by the temporal progression of interaction rather than any single piece of decisive evidence. This processual account connects two strands of theory that are often treated separately: it extends Luhmann's view of trust as a mechanism for reducing social complexity under conditions that cannot be fully known \cite{ND-luhmann2017} by grounding that reduction in moment-to-moment interactional work, and it aligns with warranting theory \cite{ND-walther2009}, which holds that the perceived reliability of an online cue depends on how difficult it is to manipulate. Integrating these perspectives, we argue that warrants do not resolve uncertainty once and for all but instead serve as provisional inputs to a continuous calibration of ``sufficient'' confidence, situating trust within the wider HCI literature on managing risk in mediated intimacy \cite{ref14}.

Our findings further challenge the assumption that deception can be understood as a clearly bounded act, such as the presentation of false information or the moment of financial extraction \cite{ref45,ref49}. Participants' accounts indicate that many forms of deception---particularly those related to relational intent and emotional investment---unfold gradually through interaction. Rapid emotional escalation, sustained ambiguity, or delayed resource requests acquire meaning only within a broader interactional trajectory, emerging retrospectively as users reinterpret earlier moments in light of later developments. Deception thus operates as a process embedded in the pacing and alignment of interaction \cite{ref17,ref47}, not a singular violation. This processual view integrates with established interactional concepts: it resonates with Goffman's framing of social encounters as ongoing accomplishments of impression management \cite{ref24}, and with accounts of trust as something incrementally calibrated rather than granted at a single point \cite{ND-mollering2006}. Reframing deception as the progressive misalignment of an interactional trajectory connects our work to this broader literature on relational sensemaking. Some situations may involve mismatched expectations or asymmetric intentions rather than deliberate fraud, making deception less a matter of detecting falsehood and more a matter of recognizing when an interaction no longer aligns with one's expectations.

A key finding is that users' ability to recognize deception does not develop solely through individual experience but is shaped by the circulation of knowledge within the community \cite{ref9,ref41}. Participants frequently relied on sharing experiences, discussing specific cases, and collectively interpreting ambiguous interactions. Conceptually, this practice can be understood as a form of \emph{distributed sensemaking}, in which interpretive labor is shared across members rather than borne by any single user \cite{ND-weick1995}, and as an expression of \emph{social translucence}, whereby community visibility into others' experiences makes otherwise opaque risks collectively legible \cite{ND-Erickson2000}. This pattern highlights the insufficiency of existing platform mechanisms---reporting systems, account verification---for addressing the gray-area risks that dominate everyday use \cite{ref39}: emotional manipulation, ambiguous relational intent, and exaggerated self-presentation are difficult to formally define or prove. Users therefore turn to informal channels, effectively constructing a distributed system of risk recognition outside formal governance structures. Community-based knowledge sharing thus functions as a compensatory mechanism, filling the interpretive gaps left by platform design \cite{ref16,ref25}, and reframing risk recognition as a socially distributed capacity rather than an individual one.

Finally, we reflect on which of these dynamics are likely to travel beyond the Chinese context and which are situated in it. Three findings appear to be strong candidates for cross-cultural transfer: the structural unverifiability of intent that platforms create, the two-axis (means/intent) organization of deception, and the compensatory emergence of community heuristics where platform and institutional verification fall short---mechanisms that presuppose only mediated dating under uncertainty, not any particular legal or cultural regime. Other findings are more clearly conditioned by the Chinese environment: the compound acquaintance-recognition and forced-outing risk attached to location visibility, the near-obligatory status of dating platforms where offline queer infrastructure is thin, the specific scam ecology (e.g., \emph{sha zhu pan} investment fraud), and the intergenerational transmission channels shaped by a semi-closeted community. We therefore read our typology not as a universal inventory but as one situated instantiation of a more general process, and we join recent queer-HCI calls \cite{ND-cui2022,ND-park2024,ND-park2025,ref42} in inviting comparative studies: examining how users in other legal, cultural, and platform regimes define, detect, and collectivize deception would test which mechanisms are structural to mediated intimacy and which are products of a particular social order---an agenda to which this study contributes the Chinese gay-community case.

\subsection{Implications for Designing for Ambiguous and Evolving Risk}

Most platform safety mechanisms today assume that risk is identifiable, reportable, and rule-bound---a user violates a policy, another user reports it, and the platform adjudicates. But the risks our participants described rarely work that way. Pian pao does not violate any written rule. Emotional manipulation leaves no obvious evidence. A marketing account chatting about daily life looks, on the surface, like any other user. If platforms want to address these gray-area risks, they need to rethink what ``safety'' means in intimate interaction contexts. Conceptually, the cases above share a common structure: risk is not a discrete event to be caught but an unfolding interpretive process, what we frame as \emph{relational risk-as-sensemaking}---an ongoing, situated judgment users assemble from accumulating cues rather than a fixed property of a profile or message. This integrative lens ties together the otherwise separate design directions we discuss below, each of which supports a different moment in that interpretive process \cite{ND-weick1995}.

One direction is to help users make sense of interactions as they unfold, rather than only verifying profiles at the point of entry. Our participants already track interactional pace, emotional escalation, and behavioral consistency over time---platforms could support this by surfacing relevant contextual cues, such as how long an account has been active, how communication patterns have shifted, or how quickly intimacy language has appeared relative to the duration of contact. The goal is not to automate suspicion but to make the information users already seek more accessible.

Another direction concerns reporting itself. Current systems typically ask users to name a specific violation, which filters out exactly the ambiguous experiences that dominate our data. A mechanism that allows users to flag an interaction as ``uncertain'' or ``uncomfortable''---without requiring them to prove wrongdoing---could better capture the texture of everyday risk and feed into platform-level pattern detection \cite{ref31,ref42}.

Our findings also underscore that community knowledge is doing much of the safety work that platforms are not. Users already share screenshots, compare notes, and distill recurring patterns into practical advice. Platforms could support this by providing structured spaces for anonymized experience sharing or connecting newer users with more experienced ones---though any such feature would need to avoid over-formalizing what currently works precisely because it is organic and peer-driven. Read through the lens of risk-as-sensemaking, these three directions are complementary rather than alternative: contextual cues enrich the raw material for interpretation, ambiguous reporting legitimizes interpretation under uncertainty, and community knowledge distributes interpretation across users over time.

All of these considerations take on additional weight in the context of gay men in China, where social stigma narrows the available channels for meeting partners and the semi-closeted status of many users amplifies vulnerability to exploitation. Design interventions developed without attention to these structural conditions risk being irrelevant at best and harmful at worst. The point is not to replace users' judgment but to better support the situated, ongoing work they are already doing to keep themselves and each other safe.

\section{Critical Reflections and Future Work}
We acknowledge several limitations of this work. Our 22 participants were recruited via snowball sampling, which likely skews toward users with stronger social ties and greater willingness to discuss sensitive topics. Those who are more isolated, or whose encounters with deception were particularly distressing, may not be represented here. Our participants also spanned a wide age range, and although this breadth let us hear across generational experiences of dating and disclosure, it also means that age-specific dynamics---differing platform fluency, distinct histories of stigma, and varied expectations about intimacy---are likely flattened in our aggregate account and warrant more targeted, cohort-sensitive investigation. In addition, all accounts were retrospective---what participants told us had already been filtered through reflection, community conversation, and the passage of time. Methods that sit closer to lived experience, such as diary studies or experience sampling, could surface dynamics that retrospective interviews miss.

Our scope was also limited in important ways. We spoke only with gay men; the landscape of deception almost certainly looks different for lesbian, bisexual, and transgender users, whose relational norms, platform choices, and exposure to risk are shaped by distinct social pressures \cite{ref25}. We also did not probe the downsides of community-based risk knowledge---the possibility that shared heuristics harden into stereotypes, that collective vigilance tips into excessive suspicion, or that informal rules end up policing who counts as a legitimate community member. These are real tensions that deserve sustained attention.

This study is rooted in the Chinese context, where legal ambiguity around homosexuality, persistent social stigma, and a particular configuration of dating platforms together shape both the kinds of deception users face and the resources they can draw on \cite{ref30,ref54}. Rather than treating this situatedness as a constraint on generalizability, we regard it as a core strength of the work: by foregrounding a marginalized community navigating a specific regulatory and cultural configuration, our findings make visible the everyday, community-grounded forms of protective work that universalizing accounts of dating-platform deception routinely overlook \cite{ND-costanza2018}. Whether the patterns we describe---especially the reliance on community knowledge as a substitute for formal platform protections---hold in contexts with different regulatory and cultural conditions remains an open question for comparative research.

\section{Conclusion}
This study has examined how gay male users in China experience, interpret, and collectively respond to deception on dating applications. Through semi-structured interviews with 22 participants, we traced the path from the interactional structures that render deception possible and difficult to detect, through the graduated judgment practices users develop in response, to the community-based processes through which risk recognition capacities are collectively generated and sustained.

Our findings challenge several prevailing assumptions in the study of online deception. First, deception on dating platforms is not primarily a matter of false information but of intentional misalignment---the gap between expressed and actual relational intent---which unfolds processually through interaction rather than residing in static profile claims. Second, trust assessment does not operate as binary verification but as ongoing uncertainty management, in which users continuously integrate multiple signals into provisional, revisable judgments of ``sufficient credibility.'' Third, the capacity to recognize and navigate deception is not an innate individual skill but a socially generated competence, produced through the circulation, abstraction, and codification of interactional experiences within community networks.

These findings contribute to CSCW scholarship by demonstrating that community-based knowledge sharing functions as a compensatory mechanism for the interpretive gaps left by formal platform governance, and that the gray-area risks characteristic of intimate interaction---emotional manipulation, ambiguous intent, relational misrepresentation---require design approaches that move beyond clear-cut violation frameworks. Rather than replacing users' judgment, platforms can better support the situated, evolving processes through which users collectively interpret and respond to risk.

When platforms cannot adequately address the ambiguous risks inherent in intimate relational contexts, users---particularly those in marginalized communities---construct their own provisional but effective infrastructure of trust and protection through community collaboration. Recognizing, supporting, and learning from these grassroots practices represents a promising direction for designing safer and more equitable dating platforms.

\bibliographystyle{ACM-Reference-Format}
\bibliography{references}

\appendix

\section{Participant Demographics}

\begin{table}[h]
\caption{Participant Demographics}
\centering
\begin{tabular}{@{}llll@{}}
\toprule
U/R & Education & Occupation & Dating APP \\
\midrule
R & Bachelor & Student & Aloha, Blued, Fanka, Soul \\
R & Bachelor & Student & Aloha, Blued \\
R & High School & Student & Aloha, Blued, Fanka \\
U & High School & Student & Blued, Soul \\
R & Master & Nurse & Blued, Fanka, Soul \\
U & Master & Teacher & Aloha, Blued, Fanka, Soul \\
R & High School & Delivery & Aloha, Blued, Fanka, Soul \\
U & Junior High & Chef & Aloha \\
R & PhD & Student & Blued, Fanka \\
R & Primary School & Farmer & Blued \\
U & Master & Editor & Soul \\
R & High School & Cleaner & Aloha, Soul \\
U & Master & Reporter & Fanka, Soul \\
U & Bachelor & Secretary & Blued, Fanka, Soul \\
R & Master & Research Asst. & Aloha, Blued, Fanka, Soul \\
R & Merely Literate & Chef & Aloha, Blued \\
U & Primary School & Labour & Blued, Soul \\
R & Merely Literate & Farmer & Aloha, Blued \\
R & PhD & Professor & Aloha, Blued \\
U & Master & Teacher & Fanka, Soul \\
R & Merely Literate & Unemployed & Aloha, Blued, Soul \\
U & High School & Student & Blued, Fanka, Soul \\
\bottomrule
\end{tabular}
\end{table}

\section{Interview Guide}

\begin{table}[h]
\caption{Semi-Structured Interview Guide}
\centering
\begin{tabular}{@{}p{0.18\linewidth}p{0.4\linewidth}p{0.36\linewidth}@{}}
\toprule
Theme & Core Question & Possible Probes \\
\midrule
Platform Use & Which dating or social apps do you usually use? How do you typically use them? & How long/often do you use them? What do you usually do on these platforms? \\
Initial Motivation & When you first started using these apps, what were you hoping to find? & Have your expectations changed over time? \\
Platform Environment & How would you describe the interaction environment on these platforms? & Similar to or different from offline social interaction? \\
Uncomfortable Experiences & Have you had interactions that made you feel uncomfortable or that later seemed problematic? & Could you describe a specific example? \\
Recognizing Problems & When did you begin to feel that something might not be right? & Triggered by something specific, or a general feeling? \\
Reflection & Looking back, how do you now interpret that experience? & Misunderstanding, misrepresentation, or something else? \\
Understanding Deception & What kinds of situations could be considered misleading or deceptive? & Must deception involve explicit lies, or also misleading intentions? \\
Patterns & What kinds of problems occur most often? & Do they happen in similar ways or follow certain patterns? \\
Motivations & Why do you think some people engage in these behaviors? & What are they usually trying to achieve? \\
Trustworthiness & How do you decide whether someone new seems trustworthy? & Particular signals or behaviors you attend to? \\
Interaction Dynamics & Does how someone communicates affect your judgment? & E.g., pace of escalation, topics brought up. \\
Consistency & Do you attend to whether what someone says is consistent over time? & Example of something that raised doubt? \\
Managing Uncertainty & When uncertain about someone, what do you usually do? & Continue observing, ask questions, slow down? \\
Verifying Identity & Do you try to verify information about someone new? & What kinds of information do you check? \\
Photo Interpretation & How do profile photos influence your judgment? & Signs that photos are authentic or representative? \\
Real-Time Confirmation & Do you confirm identity through voice/video calls or additional photos? & Under what circumstances? \\
Risk Boundaries & Are there signals that make you particularly cautious? & What behaviors or requests raise concerns? \\
Handling Risk & When you sense risk or deception, how do you respond? & Stop responding, confront, or observe further? \\
Financial Situations & If someone mentions financial difficulties or asks for money, how do you react? & Personal rules you follow? \\
Sharing Experiences & After a problematic interaction, do you talk about it with others? & Who do you talk to? \\
Collective Judgment & Do you ask others for opinions when unsure about someone? & E.g., sharing screenshots of conversations. \\
Learning from Others & Have you learned about risks from other people's experiences? & How did those stories influence you? \\
Recognizing Patterns & Have you noticed recurring patterns in problematic interactions? & Certain behaviors that appear repeatedly? \\
Community Knowledge & Are there shared ``rules'' within the community about avoiding problems? & Examples of such advice or practices? \\
Learning Process & Was your ability to judge risks different when you first started? & How did your understanding change? \\
Advice to New Users & What advice would you give someone new to these platforms? & Lessons from your own experience? \\
\bottomrule
\end{tabular}
\end{table}

\end{document}